# Superconducting Behavior of Interfaces in Graphite: Transport Measurements of Micro-constrictions

S. DUSARI, J. BARZOLA-QUIQUIA AND P. ESQUINAZI*

Division of Superconductivity and Magnetism,
Linnéstrasse 5, University of Leipzig
D-04013 Leipzig, Germany

We have studied the magnetoresistance (MR) of thin highly oriented pyrolytic graphite mesoscopic samples without and with micro-constrictions of different widths between the voltage electrodes. The MR for fields parallel to the c-axis shows an anomalous hysteresis loop compatible with the behavior expected for granular superconductors. The smaller the constriction width the larger is the anomalous hysteresis and the higher the temperature for its observation. Our results support the existence of granular superconductivity probably embedded at interfaces between crystalline graphite regions.

1. INTRODUCTION

The possibility of having some kind of superconductivity in pure graphite has been speculated years ago based on magnetization and transport data in bulk samples [1,2]. A well reproducible phenomenon in highly oriented bulk graphite samples that might be related to superconductivity is the magnetic-field driven metal-insulator transition (MIT) [3-7]. This "transition" is remarkable because the resistance at low temperatures increases by two or more orders of magnitude after applying a magnetic field of a few kOe [2]. Furthermore, this transition depends only on the field component perpendicular to the graphene planes [8], indicating its two-dimensional character. However, the origin of the MIT in graphite remains controversial [2,6,7]. Recently, high-resolution MR measurements on thin mesoscopic graphite samples showed the existence of anomalous hysteresis loops at $T \leq 25$ K supporting the existence of granular superconductivity in some unknown parts of the samples [9].

Transport measurements done in graphite samples of different thickness indicate that the MIT vanishes for samples thinner than ~50 nm suggesting that certain regions parallel to the graphene planes might be the origin of the MIT [10]. Transmission electron microscopy studies revealed that the highly oriented pyrolytic graphite samples are not homogeneous but consist on stacks of ~ 50 … 100 nm thick (c-axis direction) and micrometer long single crystalline regions [10]. The increase of the absolute resistivity, the vanishing of the metallic-like temperature dependence as well as of the magnetic field driven MIT by decreasing the thickness of the oriented graphite samples suggest that internal interfaces between crystalline regions influence the transport properties [10]. If superconductivity plays a role in the anomalous properties observed in oriented graphite samples then these interfaces may contain the superconducting regions. We note that a similar effect was found in the semi-metal Bi, a twin of graphite, where the interfaces between crystalline regions show superconducting properties up to critical temperatures of the order of 20K [11]. A possible reason for triggering superconductivity at temperatures $T > 10$K at internal interfaces within a graphite structure might be related to the increase in carrier density of the graphene sheets due to disorder and/or hydrogen influence [12].

Due to the granular nature of the presumed superconductivity (no zero resistance was ever measured yet), transport measurements directly at or of a single interface or superconducting "grain" are extraordinarily tough. Recently, the magnetic field and temperature dependence of the resistivity of several micrometers long and heterogeneously thick graphite sample was measured locating the voltage electrodes nearer one of those internal interfaces [13]. These MR results show a superconducting-like transition at low temperatures supporting the idea on the existence of superconducting regions inside graphite interfaces [13].

In this study we used a novel approach based on high-resolution MR measurements of thin mesoscopic graphite samples with micro-constrictions. We have measured clear hysteresis loops in the MR as a function of magnetic field applied perpendicular to the graphene planes. These loops are anomalous and reflect the behavior expected for granular superconductivity [14,9]. As shown in Ref.[15] the smaller the constriction width W the larger is the potential drop around the constriction region leaving the rest of the sample practically at constant potential. This expected electrostatic behavior is used in our study to increase the sensitivity of the voltage measurement mostly in the constriction region. We expect therefore that if in the constriction region similar superconducting paths exist as in other parts of the sample, an increase in the sensitivity of the measured voltage drop to the superconducting effects will be achieved. In fact, in this work we observed that the smaller the constriction width the clearer the hysteresis in the MR and the higher is the temperature where they can be measured. For the smaller constriction the anomalous hysteresis vanishes at temperatures near the temperature where the zero field resistance R(T) shows a maximum. The overall results support the existence of superconductivity at high temperatures at internal regions of highly oriented graphite. They also indicate that the metallic-like behavior of the resistivity in oriented bulk graphite sample is not intrinsic and not related to scattering of conduction electrons with phonons.

2. EXPERIMENTAL DETAILS AND SAMPLE PREPARATION

Graphite flakes were obtained by exfoliation of highly oriented pyrolytic graphite (HOPG) sample of ZYA grade (0.4° rocking curve width) and using ultrasonic technique. We selected the flakes using microscopic and micro-Raman techniques. The samples are typically 10 microns size and between 10nm and 50nm thickness with crystallographic orientation always with the main area parallel to the graphene layers. Two samples were selected for the experiments. The size of the samples are (distance between voltage electrodes x width x thickness) for sample 1 (2) 13 x 16 x 0.015 (2.6 x 6 x 0.040) $\mu m^3$. The selection of the samples has been done taking into account the differences in the temperature dependence of the resistance, as can be seen in Fig.1. Sample 1 shows a clear maximum in the resistance at T ~ 55 K, which suggests that superconducting fluctuations could play a role below it, whereas sample 2 has a much weaker maximum at lower T. Therefore, we expect to see an anomalous behavior in the MR of sample 1 below 50K and at lower temperature in sample 2.

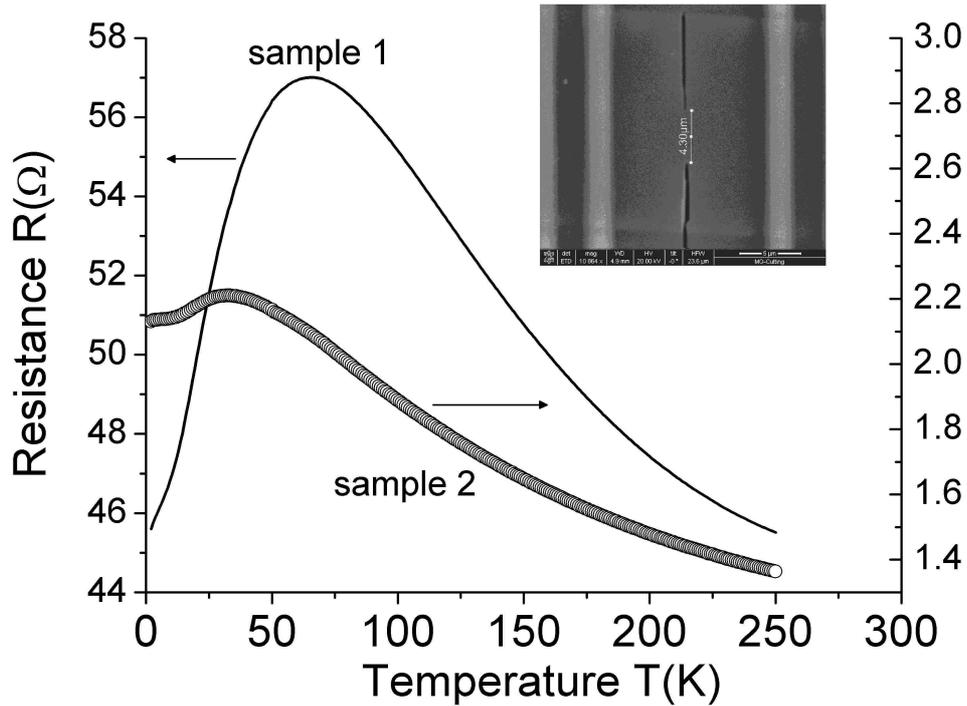

Fig.1: Resistance vs. Temperature for samples 1 and 2 (right y-axis) without constrictions and at zero applied fields. The observed temperature dependence remains for all constrictions widths. The inset shows a scanning electron microscope picture of sample 1 with a constriction width of 4.3 µm between the two voltage electrodes. The scale bar is 5 µm.

The samples were fixed on Si substrates, which have a 300 nm thick insulating $Si_3N_4$ layer on top. These substrates were fixed with silver paste on a chip carrier. The resistance measurements were done using an ac bridge (Linear Research LR700) with a relative resolution better than $10^{-5}$. The magnetic field was monitored by a small Hall sensor positioned near the sample. All the hysteresis measurements were done in field cooled (FC) state and a good reproducibility of the hysteresis details was achieved. The samples' contacts were prepared by electron-beam lithography. A bottom layer of Pt and a top layer of Au were deposited by thermal evaporation and served as contacts, see inset in Fig. 1.

The constrictions have been done using a focused ion beam (FIB) on the graphite samples previously covered with a 300 nm thick resin in order to avoid any change in the intrinsic properties of the sample during the cutting procedure [16]. The advantages of this resist are that it allows us to do patterning by electron beam lithography (EBL) in the desired shape, it is sufficiently robust in the temperature range used and it is a very bad electrical conductor. Note that the covering of the samples with the photoresist protected them also from electron-irradiation damages the scanning electron microscope produces. The penetration depth of the Ga+ ions in this resist is small enough that no Ga ions penetrates into the graphite sample outside the cutting region with exception of a thin layer of less than 1 nm at its cut-edges due to beam straggling.

## 2. RESULTS AND DISCUSSION

### 2.1 Hysteresis in the magnetoresistance

One of the most evident hints for superconductivity when neither the Meissner effect nor a zero resistant state can be measured, is the existence of a field hysteresis loop in the MR. This evidence reveals that part of the measured MR depends on pinned magnetic flux (or magnetic domains through their walls in case of ferromagnets). Figure 2 shows a typical hysteresis loop measured in sample 1 starting the sweep from +0.1T in a field cooled (FC) state. It is clearly seen that the hysteresis is anomalous in the sense that the two minima are located at the same quadrant field from which one starts the sweeping, in clear contrast to the usual hysteresis loops observed in bulk superconductors as well as in ferromagnets, where the minima (or maxima) are located at different field quadrants. This anomalous hysteresis has been already reported in conventional [14] and high-temperature granular superconductors [14,17,18] as well as in thin graphite flakes [9].

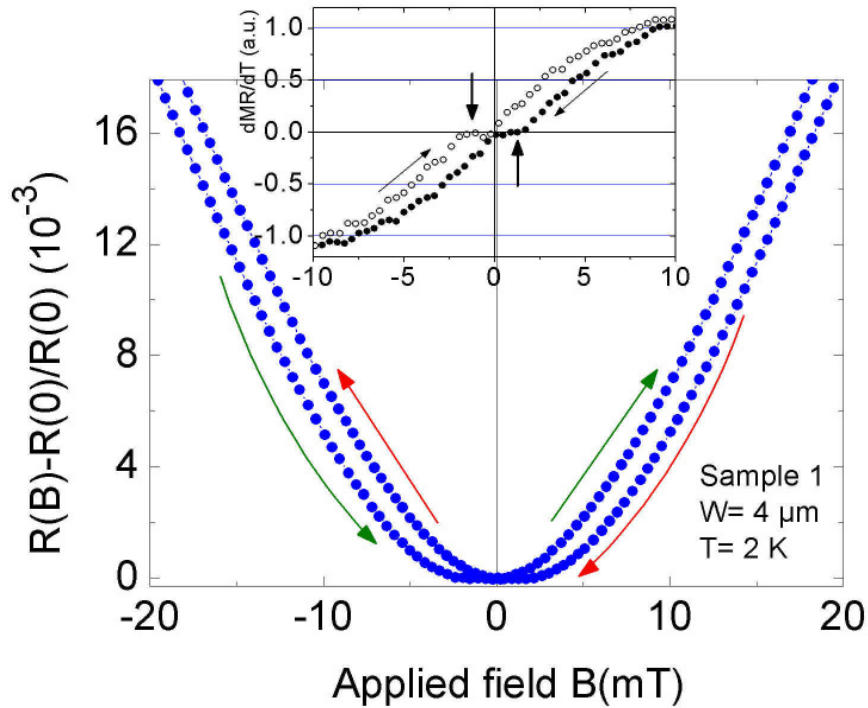

Fig.2: Magnetoresistance vs. applied magnetic field for sample 1 with a constriction 4µm size and at 2K. The input current was 1µA. Note the measured hysteresis when the field is swept from $B_{max} = \pm 100$ mT. The inset shows the first derivative of the MR vs. applied field for the two field branches as shown by the arrows. The vertical arrows indicate the position of the field $B_{min}(T)$ where the derivative of the MR vanishes.

An explanation for this anomalous hysteresis is based on a two-level critical-state model [14] where pinned fluxons exist inside the Josephson-coupled superconducting grains but also between them. These last ones are usually much less pinned and therefore can influence strongly the MR behavior, especially at low enough magnetic fields [14].

It is interesting to note that the minimum dip of the MR shown in Fig. 2 is rather flat, clearly recognized in the first derivative of the hysteresis curve, see inset in Fig. 2. This flattening has been observed before and interpreted as due to the non uniform superconducting grain size [14]. Certainly this may play a role in our case. On the other hand we note that this flattening can be well explained assuming the contribution of two MRs *in parallel*, namely, one due to the fluxon movement, i.e. the superconducting part, which for one field direction can be roughly approximated by $R_f(B) \sim a \, (abs(B - B_{min}))^{\alpha_1}$, and the second due to the intrinsic MR of the sample without superconducting grains, namely $R_i(B) \sim b \, abs(B)^{\alpha_2}$. At low fields one gets $\alpha_1 \sim \alpha_2 \sim 1.5 \ldots 2.0$.

We have also studied the dependence of the hysteresis width and its minima with the maximum field $B_{max}$ applied on FC state. Similar to the results shown in Ref.[14] for granular YBaCuO HTSC, the hysteresis minimum in our samples as well as the hysteresis width decreases the smaller $B_{max}$. At $B_{max} = 0.01$T or smaller we could not measured any hysteresis loop within experimental resolution.

2.2 Temperature and constriction-width dependence of the hysteresis loop

If the observed hysteresis loop in the MR is related to superconductivity then we expect that it should decrease with temperature and vanishes above the temperature of the resistance maximum, see Fig.1. We characterize the hysteresis loop by the difference in the MR curves between the two field directions at a certain fixed field and by the field $B_{min}(T)$ where the derivative of the MR vanishes, see inset in Fig. 2. Figure 3 shows both characterization parameters vs. temperature for sample 1 without and with two constriction widths. Whereas for the sample without constriction the hysteresis is observed only below $T \sim 5$ K, it can be observed at higher temperatures up to $T \sim 50$K the smaller the constriction width, see Fig. 3, i.e. near the temperature where R(T) shows a maximum, see Fig.1.

Sample 2 shows a shallow maximum in R(T) at $T \sim 30$ K and therefore, if our interpretation is correct, we expect to see a weaker hysteresis in the MR at lower temperatures for similar constriction widths. Figure 4 shows the MR results of sample 2 with a constriction width of 3 µm at 2K and 10K. A weaker, anomalous hysteresis in the MR can be observed only at temperatures below 10 K, in agreement with the expectations. The results of sample 2 rule out an artifact in the measurements as origin of the hysteresis and indicate that the anomalous behavior in the MR correlates with the temperature dependence of the resistance. Therefore the observed metallic-like behavior in R(T) below a certain temperature in our thin films samples is not due to electron-phonon scattering (in agreement with the discussion given in Ref.[19]), but due to superconducting fluctuations, supporting the interpretation proposed earlier [3,1].

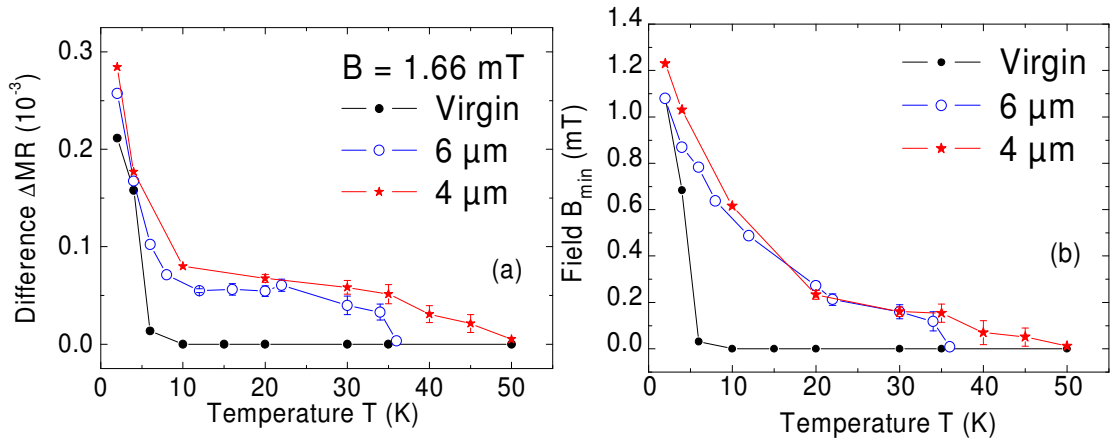

Fig.3: (a) The difference between the two field branches of the hysteresis loop in the MR vs. temperature for sample 1 in the virgin and with two different constriction widths obtained at a fixed field of 1.66mT. (b) The field $B_{min}$ vs. temperature for the same sample.

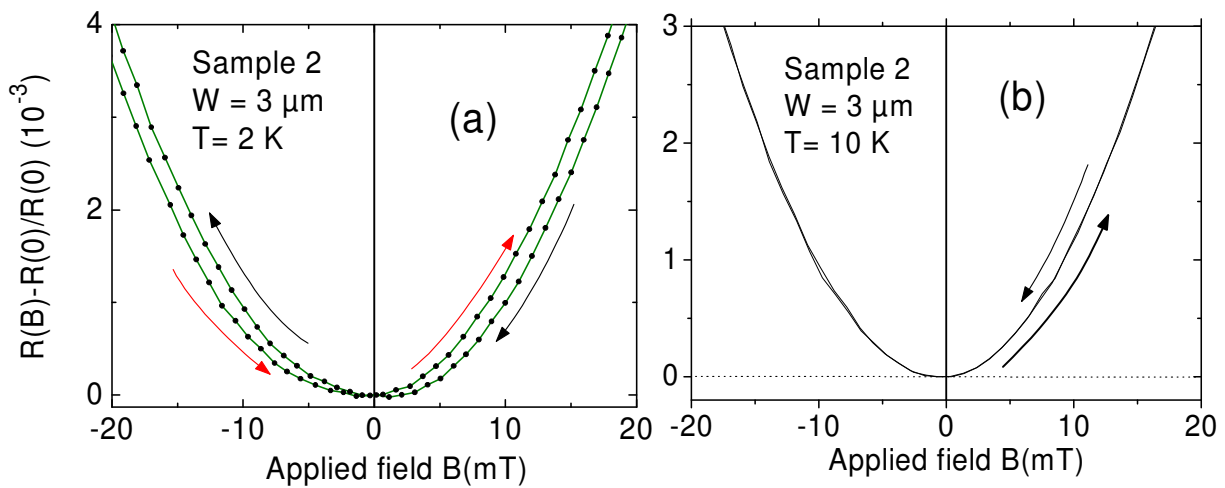

Fig.4: Magnetoresistance measured from a starting maximum field of 0.14T at the temperatures of 2K (a) and 10K (b), for sample 2 with a constriction width of 3 µm.

2.3 Simple model for the granular superconducting behavior observed in graphite

The experimental results described in Ref.[10] indicate that the superconducting regions, if they exist in graphite samples, should be localized mainly at the interfaces between crystalline regions of slightly different orientation. At these interfaces the density of carriers should be high enough to achieve high critical temperatures, provided the quasi-two dimensionality remains [12]. Taking into account transmission electron microscopy (TEM) results [10] the interfaces are distributed as depicted in Fig.5 (a) (red dashed lines) running parallel to the graphene layers and the sample surface. The measured resistance can be qualitatively treated as due to the sum of superconducting and normal in series and in parallel circuit paths. Note that the resistance of the normal paths is rather small due to the ballistic nature of the carriers at the temperature of the measurements [15]. We may expect coherent superconducting state at low enough temperatures within the superconducting paths as well as between some of them.

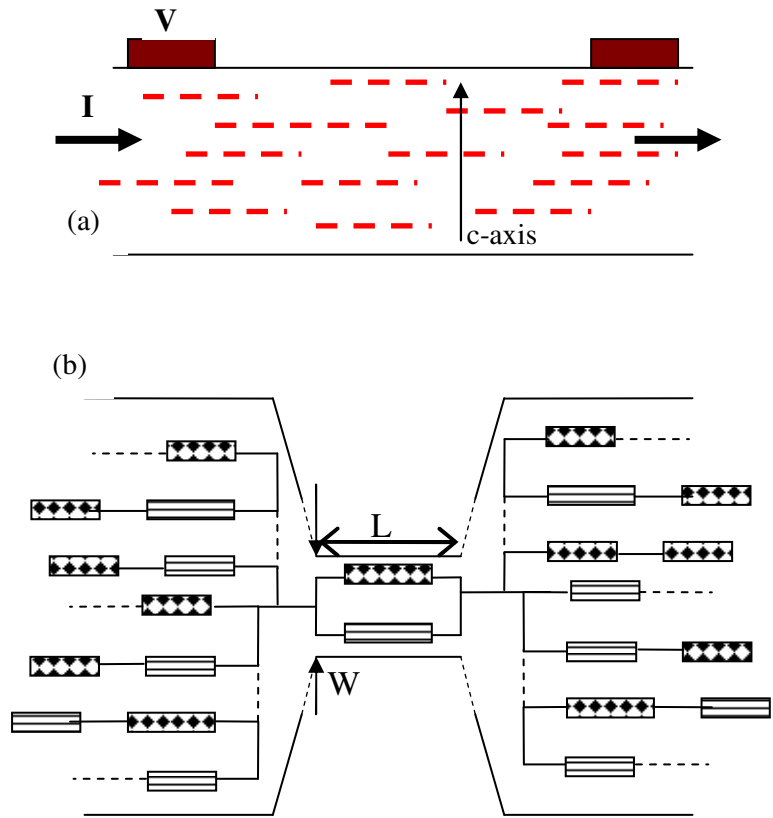

Fig.5: (a) Sketch of a graphite sample with the voltage (V) electrodes at the top surface. The c-axis is perpendicular to the graphene layers running parallel to the sample surface and to the input current (I). The dashed red lines indicate the interfaces where superconducting regions exist. This sketch is based on the TEM pictures shown in Ref.[10]. (b) Sketch of the expected average distribution of superconducting (solid diamonds) and normal (horizontal lines) resistance paths (in two dimensions) in a graphite sample with a constriction in the middle of the sample of width W and length L. In our samples L ~ 0.3 µm and W runs from 6 µm to 3 µm.

It is clear that thermal fluctuations prevent the establishment of a coherent superconducting state in parts of the sample and no zero resistance state can be achieved if the superconducting distribution is as depicted in Fig.5. In such a case and due to averaging effects a hysteresis in the MR is only observable at low enough temperatures. In case we have a constriction, which includes a smaller distribution of superconducting and normal regions, see Fig.5(b), we expect then a higher sensitivity of the voltage drop to the superconducting effects. As shown in Ref.[15] the voltage drop in a sample with constriction is expected to be concentrated mostly in the constriction region the smaller the constriction width, increasing its sensitivity to contributions of the superconducting paths.

2.4 Possible origin of superconductivity and future tasks

In Ref.[9] different possibilities were discussed as possible origin for the superconductivity in oriented graphite. The quasi-two dimensionality of the internal interfaces, the possible role of hydrogen (or defects) as carrier enhancement and the high-energy phonons of a graphene system were taken into account within the conventional mean field approach to show that the critical temperature of a hydrogen-graphene system can increase with carrier density up to ~ 60K [12]. We note that the intrinsic carrier density of ideal graphite is extremely small, as recent measurements revealed [20]. Therefore, it appears that high temperature superconductivity in graphite can be only reached if the carrier density is strongly increased without, however, losing it quasi-two dimensionality. We remark that the critical temperature of the basically three dimensional graphite-based superconductors, the so-called intercalated graphitic compounds, see e.g. Ref.[21], remain lower than 12K. Future experiments should try to increase the carrier density in a way that the coupling between graphene layers remains weak enough, as might be the case with hydrogen doping. Recently published measurements of Ca-doped graphite in the near surface region of oriented samples revealed hints for a transition at temperatures as high as 250K [22], supporting the idea that "hot-temperature superconductivity" [23] in graphite may be possible.

4. CONCLUSION

In conclusion, we have studied the behavior of the magnetoresistance with high resolution in thin, micrometer small graphite flakes. Using constrictions we could demonstrate that anomalous hysteresis loops in the magnetoresistance are measured, which survive at higher temperatures the smaller the constriction width. Our results combined with those of Refs. [9,10] support the view that certain regions in the HOPG samples as the interfaces between crystalline regions should have superconducting-like properties.

ACKNOWLEDGEMENTS: This work is supported by the DFG under DFG ES-86/16-1, S.D. is supported by the graduate school BuildMona.

REFERENCES

[1] Y. Kopelevich and P. Esquinazi, J. Low Temp. Phys. **146**, 629 (2007) and Refs. therein.

[2] Y. Kopelevich, P. Esquinazi, J. H. S. Torres, R. R. da Silva, and H. Kempa, Advances in Solid State Physics **43**, 207 (2003), ed. by B. Kramer, Springer Verlag.

[3] Y. Kopelevich *et al.*, Phys. Solid State **41**, 1959 (1999).

[4] H. Kempa, Y. Kopelevich, F. Mrowka, A. Setzer, J. H. S. Torres, R. H\"ohne, and P. Esquinazi, Solid State Commun. **115**, 539 (2000)

[5] Wang Zhi-Ming, Xing Ding-Yu, Zhang Shi-Yuan, Xu Qing-Yu, Margriet VanBael, Du You-Wei, Chinese Physics Letters **24**, 199 (2007).

[6] Xu Du, Shan-Wen Tsai, Dmitrii L. Maslov, and Arthur F. Hebard, Phys. Rev. Lett. **94**, 166601 (2005).

[7] T. Tokumoto, E. Jobiliong, E. S. Choi, Y. Oshima, and J. S. Brooks, Solid State Commun. **129**, 599 (2004).

[8] H. Kempa, H. C. Semmelhack, P. Esquinazi and Y. Kopelevich, Sol. State Commun. **125**, 1 (2003).

[9] P. Esquinazi , N. García, J. Barzola-Quiquia, P. Rödiger, K. Schindler, J.-L. Yao and M. Ziese, Phys. Rev. B **78**, 134516 (2008)

[10] J. Barzola-Quiquia, , J.-L. Yao, P. Rödiger, K. Schindler, P. Esquinazi, phys. stat. sol. (a) **205**, 2924 (2008)

[11] F. M. Muntyanu, A. Gilewski, K. Nenkov, A. Zaleski, V. Chistol, Solid State Commun. **147**, 183 (2008) and Refs. therein.

[12] N. García and P. Esquinazi, J Supercond Nov Magn **22**, 439 (2009).

[13] J. Barzola-Quiquia and P. Esquinazi, J Supercond Nov Magn **23**, 451 (2010)

[14] L. Ji, M. S. Rzchowski, N. Anand, and M. Tinkham, Phys. Rev. B **47**, 470 (1993).

[15] N. García, P. Esquinazi, J. Barzola-Quiquia, B. Ming, and D. Spoddig, Phys. Rev. B **78**, 035413 (2008).

[16] J. Barzola-Quiquia, S Dusari, G Bridoux, F Bern, A Molle and P Esquinazi, Nanotechnology **21,** 145306 (2010).

[17] Y. Kopelevich, C. dos Santos, S. Moehlecke, and A. Machado, arXiv:0108311.

[18] I. Felner, E. Galstyan, B. Lorenz, D. Cao, Y. S. Wang, Y. Y. Xue, and C. W. Chu, Phys. Rev. B **67**, 134506 (2003).

[19] See J.-P. Issi in „Graphite and Precursors", edited by P. Delhaès, Gordon and Breach science publishers, The Netherlands 2001, page 60 ff.

[20] A. Arndt, D. Spoddig, P. Esquinazi, J. Barzola-Quiquia, S. Dusari and T. Butz, Phys. Rev. B **80**, 195402 (2009).

[21] G. Csányi, P. B. Littlewood, A. H. Nevidomskyy, C. J. Pickard, B. D. Simons, Nat. Phys. **1**, 42 (2005).

[22] S.W. Han, J. D. Lee, J. P. Noh, and D.W. Jung, J. Low Temp. Phys., DOI 10.1007/s10909-010-0170-y (in press).

[23] Y. Kopelevich, P. Esquinazi, J. Torres, S. Moehlecke, J. Low Temp. Phys. **119**, 691 (2000).